\documentclass[preprint,3p,times,twocolumn]{elsarticle}
\usepackage{slashed}
\usepackage{dsfont}
\usepackage{amsmath}
\usepackage{graphicx}
\usepackage[caption=false]{subfig}
\usepackage{xspace}
\usepackage{slashed}
\usepackage{natbib}

\newcommand{\I}{\mathds{1}}

\renewcommand{\Re}{\mathrm{Re}}
\newcommand{\Tr}{\mathrm{Tr}}

\newcommand{\muhat}{\hat{\mu}}

\newcommand{\munu}{{\mu\nu}}
\newcommand{\eq}[1]{(\ref{{1}})}

\journal{Physics Letters B}
\begin{document}
\bibliographystyle{elsarticle-num}
\begin{frontmatter}
  \cortext[cor]{Corresponding author}
  \tnotetext[adp]{Report no. ADP-22-28-T1199}
\biboptions{sort&compress}

\title{Overlap quark propagator near the physical pion mass}

\author{Adam Virgili\corref{cor}} \author{Waseem Kamleh} \author{Derek B. Leinweber}
\address{Special Research Centre for the Subatomic Structure of
 	Matter, \\ Department of Physics, University of Adelaide,
 	South Australia 5005, Australia.}

\begin{abstract}

  The Landau-gauge quark propagator is calculated using overlap fermions on 2+1-flavour dynamical fermion gauge fields from the PACS-CS collaboration with pion mass $m_\pi \sim 156\text{ MeV}$ and spatial volume $\sim (3 \text{ fm})^3.$ The observed features of the mass and renormalisation functions are discussed, including a comparison with recent results using $\mathcal{O}(a)$-improved Wilson fermions on 2-flavour dynamical gauge fields.

\end{abstract}

\end{frontmatter}

\section{Introduction}

The quark propagator in momentum space offers valuable insight into the two main features of nonperturbative low-energy QCD, namely confinement and dynamical chiral symmetry breaking. 
Lattice QCD calculations can directly probe the structure of the non-perturbative quark propagator.

Previous lattice studies have utilised a range of fermion actions~\cite{Becirevic:1999rv,Becirevic:1999kb,Skullerud:2000un,Skullerud:2001aw,Oliveira:2018lln,Blossier:2010vt,Burger:2012ti,Bowman:2002bm,Parappilly:2005ei,Bowman:2005vx,Furui:2006ks,Bonnet:2002ih,Boucaud:2003dx,Kamleh:2004aw,Zhang:2003faa,Zhang:2004gv,Kamleh:2007ud,Wang:2016lsv,Pak:2015dxa,Schrock:2011hq,Burgio:2012ph}.
In principle, calculations from all valid fermion actions should agree in the continuum limit. However, at finite lattice spacing the choice of discretisation does have significant implications.

Wilson-type fermions have been used to study the quark propagator~\cite{Becirevic:1999rv,Becirevic:1999kb,Skullerud:2000un,Skullerud:2001aw,Oliveira:2018lln}, including the twisted-mass variant~\cite{Blossier:2010vt,Burger:2012ti}. However, there are associated difficulties, primarily stemming from the fact that chiral symmetry is explicitly broken by the Wilson term. This implies that the fermion propagator is no longer protected from additive mass renormalisation, and the extraction of the mass and renormalisation functions of the quark propagator becomes nontrivial.
Whilst more straight-forward `additive'~\cite{Skullerud:2000un} and `multiplicative'~\cite{Skullerud:2001aw} tree-level correction methods work well within certain regimes, they run into issues. 
A more sophisticated `hybrid' method resolves these, however at the expense of introducing ambiguities into the results~\cite{Skullerud:2001aw}.

A recent study~\cite{Oliveira:2018lln} computed the quark propagator on a large-volume lattice with dynamical $\mathcal{O}(a)$-improved Wilson fermions for the first time.
The study employs the hybrid method for the mass function and supplements it with an H4-extrapolation~\cite{Becirevic:1999uc,deSoto:2007ht} for the renormalisation function.
The nonmonotonic behaviour observed in the renormalisation function~\cite{Oliveira:2018lln} is a phenomenon previously unseen in quark propagator studies. 

Studies using staggered fermions~\cite{Bowman:2002bm,Parappilly:2005ei,Bowman:2005vx,Furui:2006ks} benefit being relatively cheap to simulate and maintaining a remnant of chiral symmetry, though the formulation is not without its own complications. In particular the fermion doubling problem is not removed but only reduced, such that a number of additional fermion species or `tastes' remain. 

The overlap fermion action~\cite{Narayanan:1993zzh,Narayanan:1993sk,Narayanan:1993ss,Narayanan:1994gw,Neuberger:1997fp,Kikukawa:1997qh} is a solution to the Ginsparg-Wilson relation~\cite{Ginsparg:1981bj}, providing an implementation of chiral symmetry on the lattice and is sensitive to the topological structures of the gauge field.
Despite its significant computational cost, the overlap action has been used extensively in lattice studies of the quark propagator~\cite{Bonnet:2002ih,Boucaud:2003dx,Kamleh:2004aw,Zhang:2003faa,Zhang:2004gv,Kamleh:2007ud,Wang:2016lsv,Pak:2015dxa}.
The advantage of the overlap lies not only in its superior chiral properties, but also the straight-forward, prescribed manner in which the mass and renormalisation functions can be extracted. The only tree-level correction necessary is to identify the kinematical momentum.

Given the different properties of the various fermion discretisations, it is of interest to compare quark propagator results obtained from the respective fermion actions.
In this work, we compute the quark propagator using the overlap fermion action on 2+1-flavour dynamical fermion gauge fields from the PACS-CS collaboration, and present the mass and renormalisation functions.

\section{Landau-gauge overlap quark propagator}

\subsection{Overlap fermions}

Within the overlap formalism~\cite{Narayanan:1993zzh,Narayanan:1993sk,Narayanan:1993ss,Narayanan:1994gw,Neuberger:1997fp,Kikukawa:1997qh}, the massless overlap Dirac operator is given by
\begin{equation}
    D_{o} = \frac{1}{2a} \left( 1 + \gamma^5 \epsilon \left(H\right) \right)\,,
\end{equation}
where $\epsilon (H)$ is the matrix sign function applied to the overlap kernel $H$.
Typically, the kernel is chosen to be the Hermitian Wilson Dirac operator, but other choices are valid and in particular the use of a kernel which incorporates smearing can have numerical advantages~\cite{Kamleh:2001ff,Bietenholz:2002ks,Kovacs:2002nz,DeGrand:2004nq,Durr:2005mq,Durr:2005ik,Bietenholz:2006fj}.
In this work, we consider the fat-link irrelevant clover (FLIC) fermion action~\cite{Zanotti:2001yb,Kamleh:2001ff,Kamleh:2004xk,Kamleh:2004aw} and choose $H=\gamma^5D_\text{flic}$ with
\begin{equation}
    D_\text{flic} = \slashed{\nabla}_\text{mfi} + \frac{a}{2} \left( \Delta^\text{fl}_\text{mfi} - \frac{1}{2}\sigma\cdot F^\text{fl}_\text{mfi} \right) + m_w \,.
\end{equation}
Here, the subscript mfi denotes the use of gauge links which have been mean-field improved~\cite{LePage:1992xa} by taking
\begin{equation}
    U_\mu(x) \to \frac{U_\mu(x)}{u_0}
\end{equation}
where
\begin{equation}
    u_0 = \langle \frac{1}{3} \Re \Tr \left[ P_\munu(x) \right] \rangle^{\frac{1}{4}}
\end{equation}
is the mean link.
The Wilson and clover terms are constructed from {\it{fat links}}, denoted by the superscript fl, which have undergone four sweeps of stout-link smearing~\cite{Morningstar:2003gk} at $\rho=0.1$.
For these terms, mean-field improvement is applied to the fat links.
The Wilson hopping parameter $\kappa$ is related to $m_w$ by
\begin{equation}
    \kappa \equiv \frac{1}{8-2am_w}\,.
\end{equation} 
The utility of the FLIC kernel is the significant improvement of the condition number following the projection of the 80 lowest-lying eigenmodes~\cite{Kamleh:2001ff}.

The massive overlap Dirac operator~\cite{Neuberger:1997bg} is defined as
\begin{equation}
    D_{o}[\mu] = (1-\mu)D_o + \mu\,,
\end{equation}
where $0 \le \mu \le 1$ is the overlap fermion mass parameter, related to the bare quark mass by
\begin{equation}
    m_q = 2m_w\,\mu\,.
\end{equation}

\subsection{Quark propagator}

The massive overlap quark propagator in coordinate space is given by
\begin{equation}
 S(x,y) = \frac{1}{2 m_{\rm w}(1-\mu)}(D_{\rm o}^{-1}[\mu](x,y)-\delta_{x,y}),
\end{equation}
where colour and spinor indices have been suppressed. The subtraction of the contact term implies that the overlap propagator satisfies
\begin{equation}
 \{ \gamma_5, S\big|_{m_q = 0 } \} = 0,
\end{equation}
mirroring the continuum chiral symmetry relation.
After taking the colour trace and transforming to momentum space, the general form of the overlap quark propagator on the lattice can be written as
\begin{equation}
    S(p) = \frac{Z(p)}{i \slashed{q} + M(p)},
    \label{eq:S_ctr}
\end{equation}
were $Z(p)$ is the renormalisation function and $M(p)$ is the mass function. Here $q_\mu$ is the kinematical lattice momentum defined by considering the tree-level propagator
\begin{equation}
    S_\text{tree}^{-1}(p)=i\slashed{q} + m_w \,,
\end{equation}
with the link variables set to unity, $U_\mu(x) = \I \ \ \forall \ \ x,\,\mu$.
This is the only tree-level correction required for the overlap quark propagator.
The simple form of Eq.~(\ref{eq:S_ctr}) is afforded by the absence of additive renormalisation in the overlap formalism.
Isolation of $M(p)$ and $Z(p)$ is straight forward.
We can rewrite Eq.~(\ref{eq:S_ctr}) as
\begin{align}
  S(p) &= \frac{-i\slashed{q}Z(p) + M(p)Z(p)}{q^2+M^2(p)} \nonumber \\
  &\equiv -i\slashed{\mathcal{C}}(p) + \mathcal{B}(p),
\end{align}
where we have defined
\begin{align}
    \mathcal{B}(p) &\equiv \frac{1}{n_sn_c}\Tr\left[S(p)\right]\, = \frac{M(p) Z(p)}{q^2+M^2(p)}\,, \\
    \mathcal{C}_\mu(p) &\equiv \frac{i}{n_sn_c}\Tr\left[\gamma_\mu S(p)\right] = \frac{q_\mu Z(p)}{q^2+M^2(p)}\,,
\end{align}
and $n_s$ and $n_c$ are the respective extents of the spin and colour indices.
Defining
\begin{equation}
    \mathcal{A}(p) \equiv \frac{q\cdot\mathcal{C}}{q^2} = \frac{Z(p)}{q^2+M^2(p)} \,,
\end{equation}
the mass and renormalisation functions are calculated with the ratios
\begin{align}
    M(p) &= \frac{\mathcal{B}(p)}{\mathcal{A}(p)}\,, \\
    Z(p) &= \frac{\mathcal{C}^2(p) + \mathcal{B}^2(p)}{\mathcal{A}(p)}\,.
\end{align}

\begin{figure*}[t]
    \centering
    {\includegraphics[width=\linewidth]{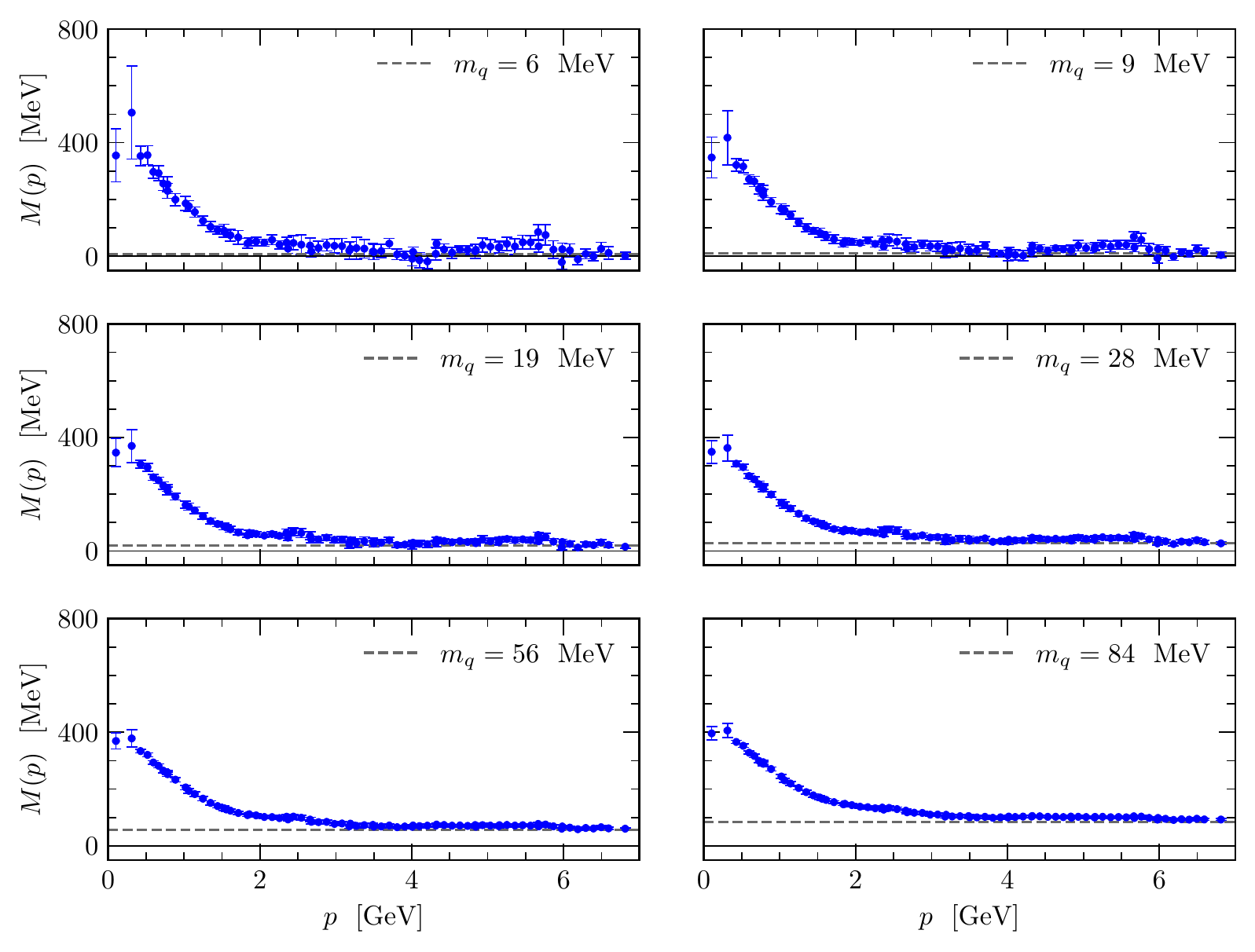}}
    \caption{Mass function $M(p)$ for all bare quark masses $m_q$ considered with $p$ on a linear scale.}
    \label{fig:M}
\end{figure*}
\subsection{Landau gauge fixing}

The quark propagator is gauge dependent, and hence requires a choice of gauge fixing condition.
In this work we use the Landau gauge condition, which in the continuum is defined by $\partial_\mu A^\mu(x) = 0.$

On the lattice, this condition is satisfied by finding the gauge transformation which maximises the $\mathcal{O}(a^2)$-improved functional~\cite{Bonnet:1999mj}
\begin{equation}
    \mathcal{F}_\text{Imp} = \frac{4}{3}\mathcal{F}_1 - \frac{1}{12u_0}\mathcal{F}_2\,,
    \label{eq:functional}
\end{equation}
where
\begin{align}
    \mathcal{F}_1 &= \sum_{x,\mu}\frac{1}{2}\Tr\left[U_\mu(x) + U^\dagger_\mu(x)\right]\,, \\
    \mathcal{F}_2 &= \sum_{x,\mu}\frac{1}{2}\Tr\left[U_\mu(x)U_\mu(x+\muhat) + U^\dagger_\mu(x+\muhat)U^\dagger_\mu(x)\right]\,.
\end{align}
The use of an improved gauge-fixing functional ensures $\mathcal{O}(a)$ improvement contained within the overlap formalism is realised in the results.
We use the Fourier accelerated conjugate gradient method~\cite{Hudspith:2014oja} to optimise Eq.~(\ref{eq:functional}).

\begin{figure*}[t]
    \centering
    {\includegraphics[width=\linewidth]{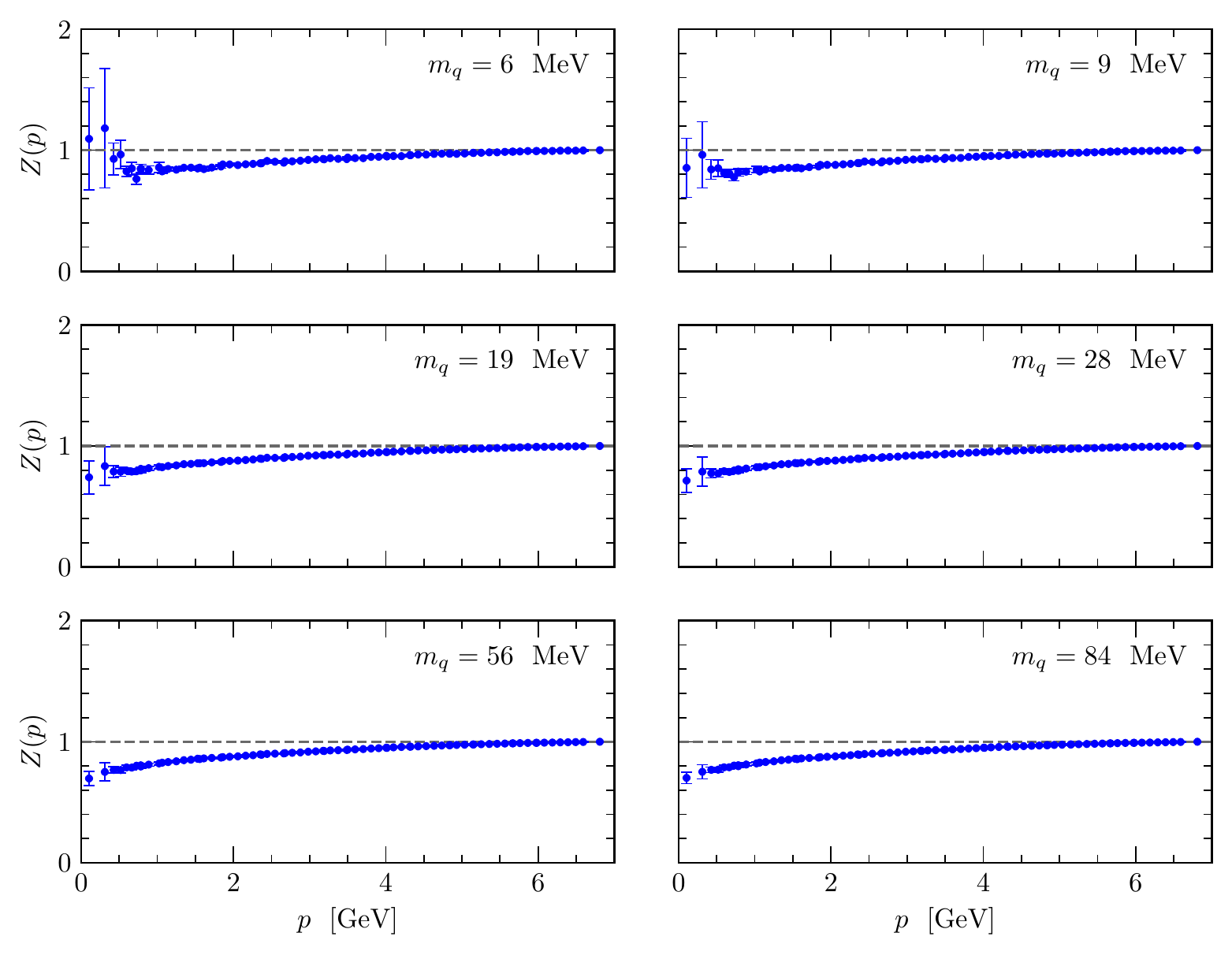}}
    \caption{Renormalisation function $Z(p)$ for all bare quark masses $m_q$ considered with $p$ on a linear scale.}
    \label{fig:Z}
\end{figure*}
\section{Results}

\subsection{Simulation parameters}

The Landau-gauge overlap quark propagator was computed on a $32^3 \times 64$ PACS-CS 2+1 flavour ensemble~\cite{PACS-CS:2008bkb} at the lightest available dynamical quark mass, corresponding to a pion mass of $m_\pi = 156$ MeV. The lattice spacing is $a=0.0933$ fm as set by the S\"ommer parameter, providing a spatial lattice volume of $\sim (3\text{ fm})^3.$
These dynamical configurations were generated using a nonperturbatively improved clover fermion action~\cite{Sheikholeslami:1985ij,CP-PACS:2005igb} and an Iwasaki gauge action~\cite{Iwasaki:1985we}.
The large volume of the configurations provides significant averaging over each configuration such that statistically accurate results are obtained on 30 gauge field configurations.

The FLIC overlap fermion action was employed at six valence quark masses $m_q = 6,\,9,\,19,\,28,\,56,\,84$ MeV where the lightest mass was tuned to match the pion mass of the ensemble.
In obtaining a favourable condition number for the inversion, the Wilson mass parameter was set to $am_w=-1.1$ corresponding to a hopping parameter of $\kappa=0.17241$ in the FLIC matrix kernel.
The matrix sign function was calculated using the Zolotarev rational polynomial approximation~\cite{Chiu:2002eh}.
The evaluation of the inner conjugate gradient was accelerated by projecting out the 80 lowest-lying eigenmodes and calculating the sign function explicitly.
Finally, a cylinder cut~\cite{Leinweber:1998im} is applied to the propagator data. 

The quark renormalisation function $Z(p)$ implicitly depends on the chosen renormalisation scale $\zeta.$ $Z(p)$ is determined by scaling the bare lattice renormalisation function such that
\begin{equation}
  Z(\zeta) = 1
\end{equation}
at the largest momentum considered, $\zeta=6.8$ GeV. The mass function is independent of $\zeta.$

\begin{figure*}[t]
    \centering
    {\includegraphics[width=\linewidth]{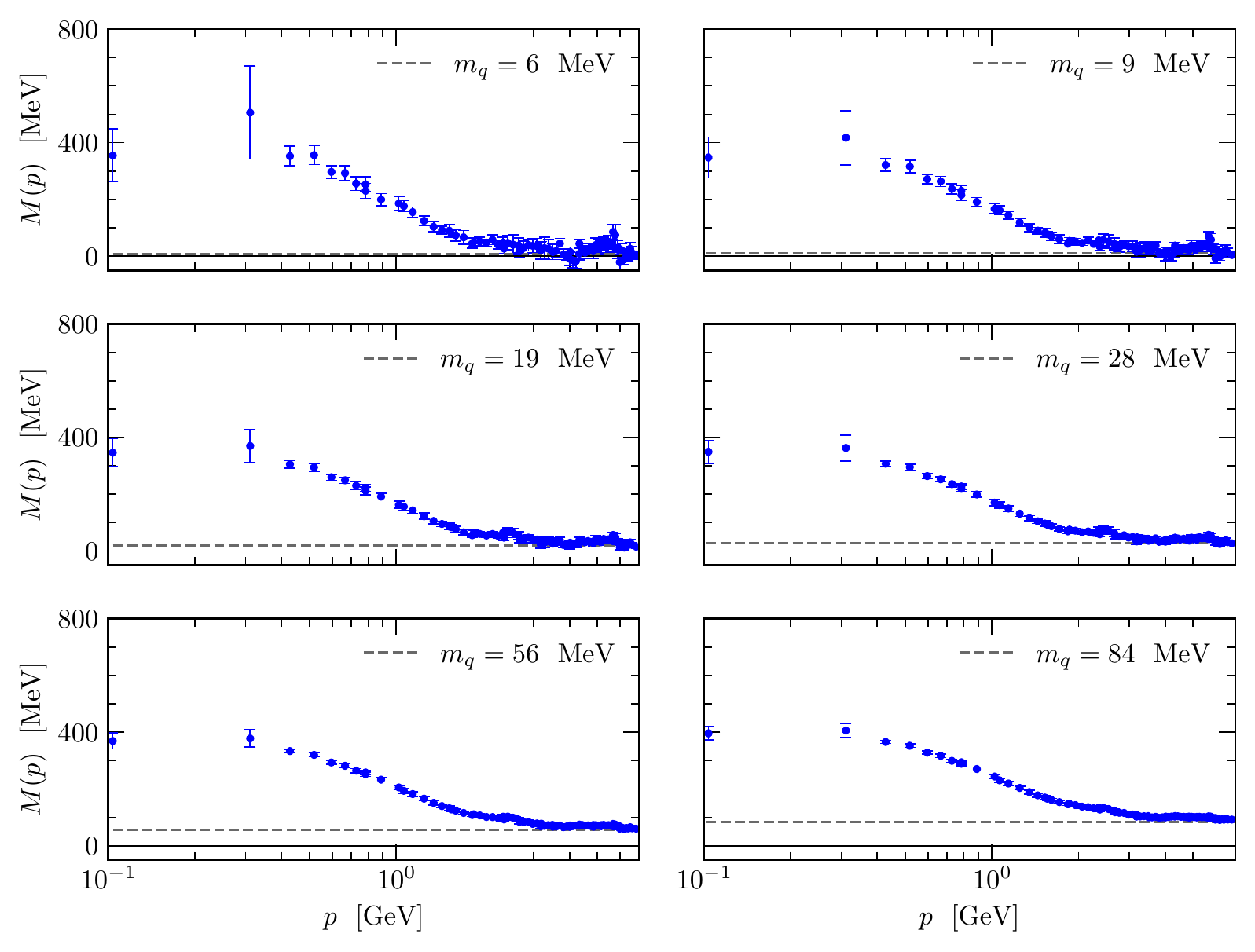}}
    \caption{Mass function $M(p)$ for all bare quark masses $m_q$ considered with $p$ on a log scale.}
    \label{fig:M-log}
\end{figure*}
\subsection{Results}

\begin{figure*}[t]
    \centering
    {\includegraphics[width=\linewidth]{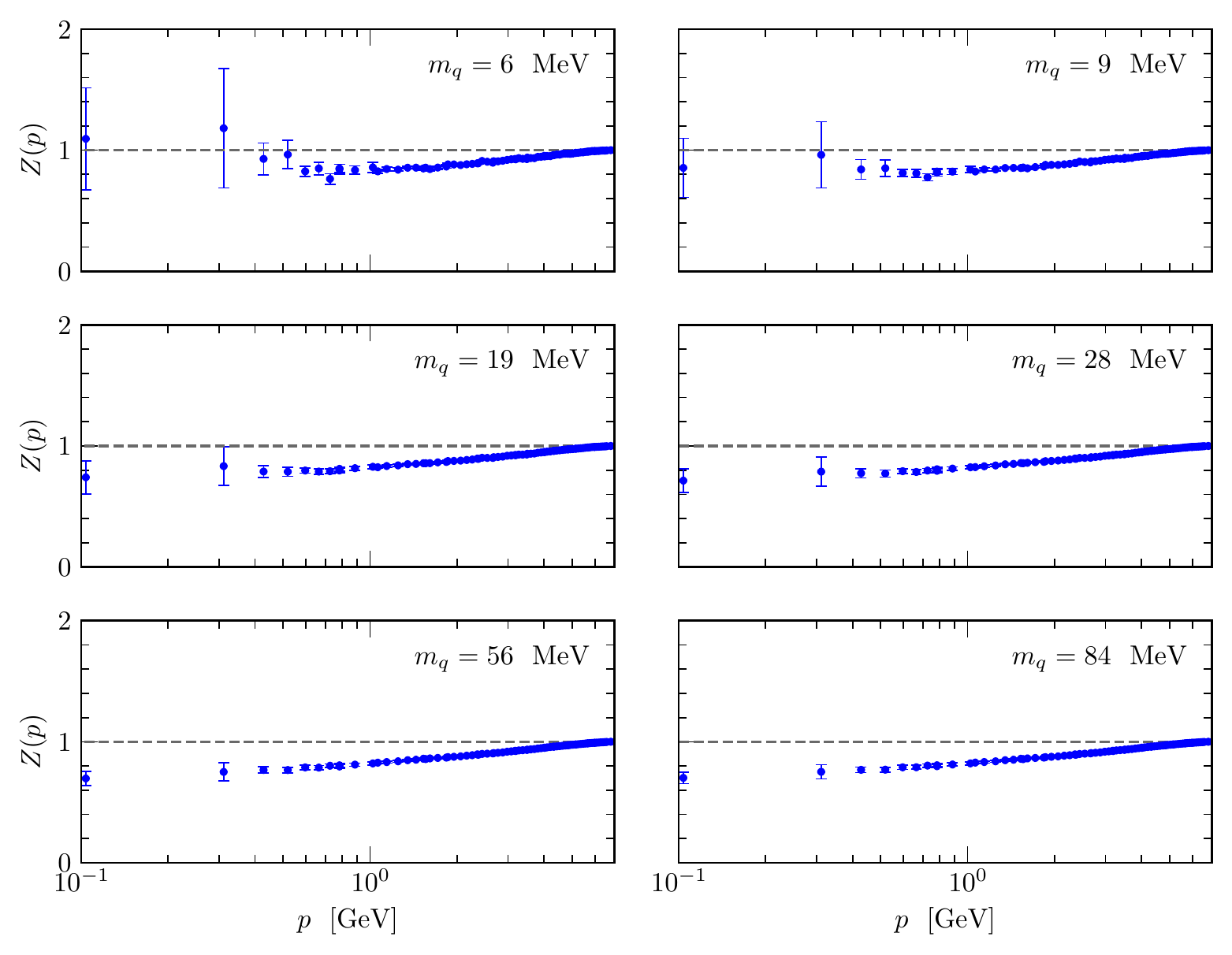}}
    \caption{Renormalisation function $Z(p)$ for all bare quark masses $m_q$ considered with $p$ on a log scale.}
    \label{fig:Z-log}
\end{figure*}

The mass and renormalisation functions $M(p)$ and $Z(p)$ for all bare quark masses considered are plotted as functions of $p$ on a linear scale in Figures~\ref{fig:M} and~\ref{fig:Z}, respectively.
To resolve the infrared behaviour  we also plot $M(p)$ and $Z(p)$ for all masses considered versus a log scale in Figures~\ref{fig:M-log} and~\ref{fig:Z-log}, respectively.

The mass function exhibits the expected qualitative features. Namely, we see agreement in the ultraviolet with the bare mass consistent with asymptotic freedom. The enhancement in the infrared that increases with decreasing bare mass is a clear signature of dynamical chiral symmetry breaking, with a generated constituent quark mass of just below 400 MeV. The steadier drop-off of the mass function away from the peak with increasing $p$ as compared to the results of Ref.~\cite{Oliveira:2018lln} is consistent with previous overlap studies. 

If we examine the mass function plotted against a log scale in Fig.~\ref{fig:M-log}, we are able to see signs of a plateau at small momenta as has been suggested elsewhere~\cite{Fischer:2003rp,Aguilar:2010cn}. The flattening of the mass function is clearer at the heavier quark masses. A future study at larger volumes in order to gain access to additional data points in the small momenta region would be of interest to confirm this behaviour.

The renormalisation functions shown in Figure~\ref{fig:Z} are consistent with the tree-level value at large momenta as expected.
Within statistical errors, $Z(p)$ is monotonically decreasing with $p$ for all bare masses considered. The log scale in Figure~\ref{fig:Z-log} enables the resolution of some fluctuations in the small momenta region at the lighter bare masses.

There is a weak mass dependence in the infrared suppression of $Z(p),$ which becomes slightly more prominent at heavier quark masses. 
This is in contrast to the findings of Ref.~\cite{Oliveira:2018lln} which found increasing infrared suppression with decreasing quark mass.  
Furthermore, Ref.~\cite{Oliveira:2018lln} observed a peak in the renormalisation function in the region of 3 GeV.
This seems to be a peculiarity of Wilson fermions, and is not seen in our results or studies using other discretisations.

\section{Conclusions}

The Landau-gauge overlap quark propagator has been calculated on a 2+1 flavour gauge ensemble with light dynamical quarks near the physical pion mass for the first time.
The signature of dynamical chiral symmetry breaking is clearly seen in the infrared enhancement of the mass function. Hints of a plateau in $M(p)$ at small momenta can be resolved when plotted on a log scale.

The behaviour of the renormalisation function is consistent with previous smaller-volume studies using overlap fermions.
The advantage of using a chiral fermion action to study the quark propagator is made clear with the observation that $Z(p)$ monotonically decreases with $p$ (up to statistical fluctuations in the far infrared at the lightest masses considered).
This is in contrast to a previous calculation~\cite{Oliveira:2018lln} that explored the Wilson fermion propagator with dynamical quarks on a large-volume lattice and found nonmonotonic behaviour in the renormalisation function.

Future investigations using even larger volume lattices would provide access to smaller nontrivial momenta, enabling a better resolution of the infrared behaviour of the mass and renormalisation functions. These results can inform theoretical formalisms that depend on knowledge of the fundamental propagators of QCD constituents~\cite{Skullerud:2002ge,Skullerud:2003qu,Bhagwat:2004kj,Fischer:2006ub,Cucchieri:2008qm,Kizilersu:2021jen}. Of course, it is desirable to seek an understanding of the non-perturbative properties of the quark propagator by examining the features of the QCD ground-state vacuum fields that give rise to these phenomena. In particular, the role of topologically-motivated degrees of freedom~\cite{Trewartha:2013qga} such as centre vortices~\cite{Trewartha:2015nna} are of contemporary interest.

\section{Acknowledgments}

We thank the PACS-CS Collaboration for making their configurations available via the International Lattice Data Grid (ILDG).
This research was undertaken with resources provided by the Pawsey Supercomputing Centre through the National Computational Merit Allocation Scheme with funding from the Australian Government and the Government of Western Australia. Additional resources were provided from the National Computational Infrastructure (NCI) supported by the Australian Government through Grant No. LE190100021 via the University of Adelaide Partner Share.
This research is supported by Australian Research Council through Grants No. DP190102215 and DP210103706.
WK is supported by the Pawsey Supercomputing Centre through the Pawsey Centre for Extreme Scale Readiness (PaCER) program.

\bibliography{Bibliography.bib}

\end{document}